# Response to "Response to Comment on 'Performance of a Spin Based Insulated Gate Field Effect Transistor'" [cond-mat/0607432]

S. Bandyopadhyay

Department of Electrical and Computer Engineering,

Virginia Commonwealth University, Richmond, VA 23284

M. Cahay

Department of Electrical and Computer Engineering and Computer Science,

University of Cincinnati, Cincinnati, OH 45221

ABSTRACT

We show that the arguments in the posting cond-mat/0607432 by Flatté and Hall are flawed and untenable. Their spin based transistor cannot work as claimed because of fundamental scientific barriers, which cannot be overcome now, or ever. Their device is not likely to work as a transistor at room temperature, let alone outperform the traditional MOSFET, as claimed.

Flatté and Hall have proposed a spin-based field effect transistor [1] that works as follows: Nearly 100% spin polarized electrons are injected from the source region into the channel. If the gate voltage is low, the spin orbit interaction in the channel is weak so that the electrons do not flip spin and arrive at the drain with their spin polarizations mostly intact. The drain is an ideal half-metallic ferromagnet magnetized *anti-parallel* to the source. It blocks these electrons from transmitting and the current is nearly zero. When the gate voltage is turned up, the spin orbit interaction in the channel increases in strength and carriers flip their spin before they reach the drain. Therefore, many carriers arriving at the drain have their spins aligned along the direction of the drain's magnetization. These are transmitted, resulting in a large drain current. Thus, by changing the gate voltage, one can modulate the source-to-drain current and realize transistor action. Flatté and Hall claim that this transistor will work at room temperature, have an on-to-off current ratio of $10^5$, and a lower threshold voltage than a MOSFET, resulting in lower energy dissipation during switching.

In [2], we questioned these claims. We believe that the Flatté-Hall device cannot even work as a proper transistor at room temperature, let alone outperform the traditional CMOS device. Flatté and Hall have recently posted a response to our objections [3]. Here we show that their response contains flawed arguments. As a result, our objections stand.

First, let us reiterate our basic objection to the Flatté-Hall device. These authors claim (see [3]) that the on-to-off current ratio in their device is $10^5$ at room temperature. We hold that this is impossible and a more realistic ratio is 10, if even that much. Such a low ratio makes the device useless in all mainstream applications. We elucidate this below.

Unless the source that injects spin-polarized electrons into the channel of the Flatté-Hall device is an ideal, 100% spin-polarized, half metallic ferromagnet, and unless there is no loss of spin polarization at the source-channel interface, the spin injection efficiency is less than 100%. Let us assume that it is $\eta$. Then,

$$\eta = \frac{I_{maj} - I_{min}}{I_{total}} \qquad (1)$$

where $I_{maj}$ is the current due to majority spins, $I_{min}$ is the current due to minority spins and $I_{total} = I_{maj} + I_{min}$. Therefore, $I_{min} = [(1-\eta)/2] I_{total}$. Assuming (generously) that the drain is a 100% efficient spin selective detector, $I_{min}$ will be the leakage current ($I_{off}$) that flows through the transistor when it is "off". When the transistor is "on", the current that flows is due to spin randomization in the channel (spin flipping). If randomization is complete, then the spin polarization of the current arriving at the drain is zero. Therefore, at best 50% of the arriving carriers will have their spins aligned along the drain. As a result, the maximum possible value of the on-current is $I_{on} = (1/2)I_{total}$. Consequently, the *maximum* on-to-off current ratio is

$$\frac{I_{on}}{I_{off}} = \frac{1}{1-\eta} \qquad (2)$$

In order for this to equal $10^5$, as claimed by Flatté and Hall, $\eta \geq 99.999\%$. Thus, even 99% spin injection efficiency is not enough! The injection efficiency at the source contact has to reach 99.999%. If we consider the fact that the drain cannot be a perfect spin detector either, then the required spin injection efficiency at the source end is even higher. Flatté and Hall do not dispute any of this, but claim in their response [3] that even though such high injection efficiency is not currently achievable, it may become achievable by 2018, because (in their view) there is no fundamental barrier to achieving 100% injection efficiency. Here, we show that there *are fundamental barriers* to ~100% spin injection efficiency, particularly at room temperature.

Flatté and Hall have proposed two routes to achieving ~100% spin injection efficiency: (1) the use of 100% spin-polarized half metals as spin injectors, and (2) the use of spin selective barriers. Unfortunately, there can be *no* half metals with 100% spin polarization at any temperature above absolute zero. Ref. [4] has shown that all half metals lose their high degree of spin polarization at temperature T > 0 K because of magnons and phonons. Even at T = 0 K, there are no ideal half metals with 100% spin polarization because of surfaces and inhomogeneities [4]. Therefore, half metals will not achieve ~100% spin injection efficiency, *ever*, even at 0 K, let alone room temperature. Consequently, half metals are not a viable route.

Spin selective barriers can at best transmit one kind of spin at one specific injection energy. The best spin selective barriers use resonant tunneling. At 0 K, the transmission energy bandwidth can approach zero, but at any non-zero temperature, thermal broadening of the carrier energy will ensure that the spin injection efficiency is far less than 100%. Therefore, this route will not work either at room temperature, not now and not ever.

In summary, there are no known methods to approach 100% spin injection efficiency because there are no half metals at T > 0 K, and there are no perfect spin selective barriers at T > 0 K. *Thus, there are fundamental barriers* to achieving ~100 % spin injection efficiency at T > 0 K. There may be other ways of achieving high spin injection efficiency in addition to the two that Flatté and Hall mention (see, for example, [5]), but they too do not work at T > 0 K. Thus, unless revolutionary new ideas for spin injection are found, there is no hope of achieving 100% spin injection efficiency at room temperature either now, or in 2018, or 3018. Simply stated, there is no strategy in sight that can produce, even theoretically, 100% spin injection efficiency at room temperature. Flatté and Hall's optimism (that somehow ~100% spin injection at room temperature will be possible by 2018) has no scientific rationale.

The largest electrical spin injection efficiency demonstrated from a permanent ferromagnet so far is ~70% [6]. We showed in [2] that this results in a current on-to-off ratio of 3.3, not $10^5$. Even if 90% spin injection efficiency is achieved at room temperature by 2018, the resulting current on-to-off ratio will be only 10. That is not sufficient for any mainstream application. Therefore, we believe that the Flatté-Hall device is not a viable transistor.

Today's CMOS transistors have a current on-to-off ratio exceeding $10^5$, which is unlikely to be ever possible for the Flatté -Hall device. This alone is sufficient to make this device non-competitive with the CMOS device and most other transistor devices currently extant.

In [2], we provided a direct comparison between the Flatté-Hall device and a MOSFET. We showed that if the *same structure* is used as either a traditional CMOS transistor, or the Flatté-Hall spin transistor, the CMOS version can always have a lower threshold voltage (and therefore win) by lowering the carrier concentration sufficiently[1]. In [3], Flatté and Hall challenge this

---

[1] Flatté and Hall also ignored the voltage drop across the gate insulator in their original calculation. They now address this issue in ref. [3]. They first state that the voltage drop across the oxide in CMOS is 70% of the total. Then they state: "if the same oxide is used for the spin-FET as planned for CMOS, the presence of the oxide changes the total spin-FET capacitance by only ~ 10%, and the voltage drop across the oxide is ~ 10% of the total voltage drop, changing the threshold voltage by ~10%." We ask: what has capacitance got to do with the voltage drop across the oxide? The total voltage drop is the sum of the drop across the oxide and the transverse drop across the channel plus substrate. This is a simple voltage divider circuit. The drop across the oxide is

conclusion by invoking an assumed 60 mV/decade sub-threshold swing 'limit' on CMOS transistors at room temperature. They evidently believe that because of this 'limit', the minimum threshold voltage required to achieve a current on-to-off ratio of $10^5$ at room temperature will be 60 x 5 = 300 mV. This line of thinking is *wrong*. The 60 mV/decade is *not* a fundamental limit. It arises only as long as carriers in the source region are in quasi-equilibrium (governed by Boltzmann statistics) and are injected into the channel by thermionic emission over the source barrier. If the carriers in the source region are "hot", they are not governed by Boltzmann statistics and the 60 mV/decade limit does not apply. Furthermore, carriers need not be injected by thermionic emission. It is well known that the gate voltage can modulate the width of the source barrier, rather than its height (or even relative band alignments as in the case of Esaki tunnel field effect transistors), to inject carriers into the channel by *tunneling*. This mode has no 60 mV/decade limit because no energy barriers are raised or lowered. In fact, the sub-threshold swing can be theoretically zero in a properly designed device with tunneling injection, in which case, the threshold voltage can also approach zero.

Having stated the above, there is, however, a *practical* lower limit on the threshold voltage imposed by noise. This limit applies to all devices, including the Flatté-Hall device and the MOSFET. Ref. [7] has shown that in modern day digital circuits, it is necessary that $V_{th}/V_n > 12$ so that bit error rates remain within acceptable limits. Here, $V_{th}$ is the threshold voltage and $V_n$ is the noise voltage given by [7]:

---

$R_{oxide}/(R_{oxide}+R_{channel+substrate})$ times the total voltage drop, where 'R' is the resistance of the relevant section. If the same oxide is used for the spin-FET and the CMOS, then $R_{oxide}$ is the same for both. If $R_{channel+substrate}$ is also same for both, then the drop across the oxide will 70%, not 10%, of the total drop even for the spin-FET. We will not belabor this point, but remonstrate that the drop across the gate insulator is not negligible unless that insulator is extremely thin. Making the insulator too thin is inadvisable since it causes gate leakage and destroys the isolation between the gate and the drain (input and output) making the transistor useless. In modern CMOS, the gate insulator is *intentionally* made thick to avoid gate leakage. This tends to decrease the gate capacitance and hence the transconductance, which is then compensated for by using high−κ dielectrics as the gate insulator.

$$V_n = \sqrt{\frac{kT}{C_g}} \qquad (3)$$

where $kT$ is the thermal energy and $C_g$ is the gate capacitance. Therefore,

$$V_{th} \geq 12\sqrt{\frac{kT}{C_g}} \qquad (4)$$

In modern integrated circuits, the clock frequency (speed) is not limited by the individual device capacitance, but rather by the interconnect- and line-capacitances. The latter are about 1 fF. Therefore, no significant speed advantage accrues from reducing the gate capacitance of a transistor to below 1 fF[2]. Assuming $C_g$ = 1 fF, there is a universal (transistor-independent) limit on the room temperature threshold voltage [in today's circuits] that can be calculated from Equation (4). It is $V_{th}$ = 25 mV. There appears to be no reason why a CMOS cannot operate with a threshold voltage as low as 25 mV at room temperature.

Flatté and Hall also contend in their response [3] that our analysis is valid at T = 0 K, while theirs is valid at T = 300 K. We have no idea what might be the basis of this contention. Perhaps it is their mistaken belief that the 60 mV/decade sub-threshold swing at room temperature will constrain any low-threshold-voltage CMOS to operate at low temperatures. Since nothing limits the sub-threshold swing to 60 mV/decade, there is no reason why a low-threshold-voltage CMOS (with a threshold voltage larger than 25 mV) cannot operate at room temperature. In fact, it is more realistic to say that our analysis is valid at arbitrary temperatures, while theirs is valid only at absolute zero, since they assume 100% spin injection efficiency.

---

[2] Because of this reason, there is really no advantage to having a low gate capacitance in the Flatté-Hall device, contrary to what they claim.

In conclusion, we find that the Flatté-Hall device is not a viable transistor, it cannot operate at room temperature with a current on-to-off ratio remotely approaching $10^5$ because of fundamental reasons, and that their response to our criticism is predicated on flawed arguments.